\documentclass[11pt]{article}
\usepackage{amsmath,amssymb,amsfonts}
\usepackage[english]{babel}
\makeatletter
\@addtoreset{equation}{section}
\makeatother

\def\nn{\nonumber} \def\bd{\begin{document}} \def\ed{\end{document}}
\def\ds{\documentstyle}
\let\bm=\bibitem
\newcommand{\be}{\begin{equation}}
\newcommand{\ee}{\end{equation}}
\newcommand{\bea}{\setlength\arraycolsep{2pt} \begin{eqnarray}}
\newcommand{\eea}{\end{eqnarray}}
\newcommand{\hoch}[1]{$\, ^{#1}$}
\def\p{\partial}
\newcommand{\auth}{
Jun-Jin Peng\hoch{\dagger}, Shuang-Qing Wu}

\begin{document}

\vspace{25pt}
\begin{center}
{\large {\bf Hawking Radiation of Black Holes in Infrared Modified Ho\v{r}ava-Lifshitz  Gravity}}

\vspace{10pt}
\auth

\vspace{10pt}
{\it College of Physical Science and Technology, Central China Normal
University,\\ Wuhan, Hubei 430079, People's Republic of China}

\vspace{20pt}
E-Mail: pengjjph@163.com\hoch{\dagger}
\vspace{20pt}

\textbf{Abstract}
\end{center}
We study the Hawking radiation of the spherically symmetric, asymptotically
flat black holes in the infrared modified Ho\v{r}ava-Lifshitz gravity by
applying the methods of covariant anomaly cancellation and effective action,
as well as the approach of Damour-Ruffini-Sannan's. These black holes behave
as the usual Schwarzschild ones of the general relativity when the radial
distance is very large. We also extend the method of covariant anomaly
cancellation to derive the Hawking temperature of the spherically symmetric,
asymptotically AdS black holes that represent the analogues of the
Schwarzschild AdS ones.

\voffset=-.90pt
\vspace{40pt}

\section{Introduction}\label{one}
Recently, Ho\v{r}ava proposed a power counting renormalizable theory for the
(3+1)-dimensional quantum gravity \cite{Horava}, motivated by the Lifshitz
model in condensed matter physics. This theory, called as Ho\v{r}ava-Lifshitz (HL)
gravity, is believed to be the potential ultraviolet (UV) completion of
general relativity. In the infrared (IR) limit (setting the dynamical coupling constant
$\lambda=1$ in the action), it recovers general relativity. Some important
characters of the HL gravity are that the Lorentz invariance
is violated at the UV level and the scaling between time and space is also
anisotropic.
The latter indicates that this model is not invariant under the full diffeomorphism
group of general relativity but has the foliation preserving diffeomorphism
invariance for generic values of $\lambda$.
Inspired by the HL gravity, a great number of
works have appeared to investigate the classical solutions
\cite{Kehagias,LuMeiPope,MIPark,Solutions}, the black hole physics
\cite{CiteKeha,ACast,BHphy} and other aspects \cite{quafiled,gwave,cosmolo}.

In order to reduce the number of the independent coupling constants, in
\cite{Horava}, the detailed balance condition was imposed. One result from
the imposition is that the generic IR vacuum of the HL
gravity is not a Minkowski vacuum but the one in anti-de Sitter background.
Besides, in the IR limit, both the Newton constant and the speed of light
are dependent on the cosmological constant. For the aim to make the
HL gravity recover the Minkowski vacuum in the IR limit,
the theory was modified by adding a term ``$\mu^4R$~" \cite{Kehagias} to
the original action in \cite{Horava}. Such a modification softly breaks
the detailed balance condition and changes the IR properties. But the ones
at the UV level are not affected. In terms of the IR modified
HL gravity, a static, spherically symmetric black hole
solution was presented in \cite{Kehagias}. This black hole behaves like
the standard Schwarzschild one when the radial distance is sufficiently
large. In \cite{MIPark}, the author obtained the black hole and cosmological
solution that is very similar with the usual Schwarzschild-AdS black hole at
long distances.

On the other hand, Hawking radiation \cite{SWH} has been regarded as a very
attractive quantum phenomenon of black holes or other geometric backgrounds
with an event horizon. It may open way for understanding the quantum gravity.
Since this quantum effect was discovered more than thirty years ago, several
different approaches have been developed to derive it. One of them is the
gravitational and gauge anomaly cancellation method recently proposed by
Robinson and Wilczek (RW) \cite{RW}. They first obtained the Hawking temperature
of the Schwarzschild black hole by considering the gravitational anomaly.
Subsequently, taking into account both gravitational anomaly and gauge
anomaly, Iso, Umetsu and Wilczek extended RW's method to the charged black
hole \cite{IUW} and the rotating black hole \cite{RotatingAnom}. This anomaly
cancellation method treats the Hawking radiation as a compensation of the
quantum anomaly that breaks the classical symmetry near the horizon. In terms
of the two dimensional effective metric, which is obtained due to the fact
that an infinite collection of (1+1)-dimensional fields can effectively describe
the quantum field near the horizon of the original higher dimensional space-time,
one can derive the fluxes of charges (if there exists gauge field) and energy
momentum tensor by solving the Ward identities under proper boundary conditions
associated with the Unruh vacuum. These fluxes are identified with the Hawking
ones. Following RW's method, a lot of works have appeared to investigate the
Hawking radiation of black objects in various dimensions
\cite{MurataJ,RotAnomB,BonoraC,WuPCQG1,Sphano,Wueff,BRing,BKulk,CAstr,GangKul,Baner}.

There are two forms of anomalies. One is consistent and the other is covariant.
For the original method proposed by RW in \cite{RW,IUW}, the expressions of
chiral anomalies take the consistent forms, but the boundary condition to
fix the arbitrary parameters is covariant. In \cite{BKulk}, it was argued that
the anomaly cancellation method can be unified in terms of the single covariant
expressions, namely, the fluxes of charges and energy momentum tensor can be
obtained via covariant gauge anomaly and covariant gravitational anomaly,
together with the vanishing of the covariant current at the event horizon. Such
an argument makes the anomaly cancellation method more economical and conceptually
cleaner. Since then, on basis of the development in \cite{BKulk}, Hawking
radiation of black strings in various dimensions was discussed \cite{CAstr}. Other
extensions can be found in \cite{GangKul,Baner}.

It is worth noting that the boundary condition plays a crucial role in fixing the
Hawking fluxes for both the RW's original method and the one of the covariant
anomaly cancellation developed later. In fact, by imposing proper boundary
condition at the event horizon only, Banerjee and Kulkarni \cite{BanerKeff} made
use of the covariant current and energy momentum tensor, generated from the two
dimensional effective action \cite{HLca}, to derive the Hawking fluxes of the
spherically symmetric charged black holes. Their results are in agreement with
those calculated via the approaches of Unruh vacuum \cite{RotatingAnom} and the
trace anomaly \cite{AShirasakaT}. Some applications of the covariant effective
action method generalized by Banerjee and Kulkarni appear in \cite{effappli}. In
our work \cite{Wueff}, this method has been extended to reproduce the Hawking
fluxes of the Schwarzschild black holes in the isotropic coordinates where the
determinant of the metric vanishes at the horizon.

So far although some properties of the black holes in HL gravity
have been investigated, the study on their Hawking radiation is still vacant. In
this paper, we shall fill this gap by applying the methods of covariant anomaly
cancellation and the covariant effective action to exploit the Hawking radiation
of the spherically symmetric, asymptotically flat black holes \cite{Kehagias} in
the IR modified HL gravity. By comparison, we will also calculate
their Hawking temperature via the method of Damour-Ruffini-Sannan (DRS) \cite{DRS}.
Besides, we shall study the Hawking radiation of the spherically symmetric,
asymptotically AdS black holes \cite{MIPark} through the covariant anomaly
cancellation method. The remainder of the present paper is organized as follows.
In section \ref{two}, we shall briefly review the IR modified HL
gravity and the spherically symmetric, asymptotically flat black hole solutions in
\cite{Kehagias}. In section \ref{three}, we will use the covariant anomaly
cancellation method to derive the Hawking temperature of these black holes. By
applying the covariant effective action approach, we reproduce the same Hawking
fluxes as those obtained via the method of covariant anomaly cancellation in
section \ref{four}. In section \ref{five}, we compute the Hawking temperature of
the asymptotically flat black holes via the DRS approach. In section \ref{six},
the analysis in section \ref{three} is extended to study the Hawking radiation of
the spherically symmetric black holes with cosmological constant \cite{MIPark}.
We present our conclusions in the last section.

\section{Black Holes in Infrared Modified Ho\v{r}ava-Lifshitz  Gravity}\label{two}

Let us start with the metric ansatz in the four dimensional HL
gravitational theory. Using the ADM-like decomposition and introducing the lapse
function $N$, the shift vector $N_i$ and the spatial metric $g_{ij}$, we have
\be
ds_4^2= - \left(N^2-N^kN_k \right) dt^2 -2g_{ij} N^i dt dx^j
 +g_{ij}dx^i dx^j  \, .
\label{metric}
\ee
In what follows, the action with IR modification takes the form \cite{Kehagias}
\bea
I &=& \int dtd^3x \sqrt{g}N\left[\frac{2}{\kappa^2}(K_{ij}K^{ij} -\lambda
K^2)   +\frac{\kappa^2\mu^2\Lambda_W}{8(1-3\lambda)}
\left(R -3\Lambda_W+ \frac{1-4\lambda}{4\Lambda_W}R^2 \right)
+\mu^4R  \right.  \nn \\
&& \left. -\frac{\kappa^2}{2w^4} \left(C_{ij} -\frac{\mu w^2}{2}R_{ij}\right)
\left(C^{ij} -\frac{\mu w^2}{2}R^{ij}\right)\right] \, ,
\label{action}
\eea
where $\lambda$, $\mu$, $w$, $\kappa$ are constant parameters, the extrinsic
curvature $K_{ij}$ is defined by
\be
 K_{ij}=\frac{1}{2N}\left[\partial_tg_{ij}-2\nabla_{(i}
N_{j)}\right]\  \, ,
 \ee
and the Cotton tensor $C^{ij}$ is read off as
\be
 C^{ij}=\frac{\epsilon^{ikl}}{\sqrt{g}}\nabla_k
\left(R^{j}_{~l}-\frac{1}{4}R \delta^j_{~l}\right) \, .
 \ee
Compared with the original action in \cite{Horava}, the added term ``$\mu^4R$~"
in the action (\ref{action}) softly violates the detailed balance condition and
modifies the IR properties of the HL gravity \cite{Kehagias},
however, it does not amend the UV ones. Particularly, in the IR limit by setting
\bea
\lambda&=&1\, , \qquad \Lambda_W=0 \, ,\nn \\
c^2 &=&\frac{\kappa^2\mu^4}{2} \, , \qquad
G_N = \frac{\kappa^2}{32\pi c} \, ,
\eea
the action (\ref{action}) reduces to the standard Einstein-Hilbert one in the
ADM formalism
\be
I_{EH} = \frac{1}{16\pi G_N}\int d^4x \sqrt{g}N\left(K_{ij}K^{ij} -
K^2+R \right) \, ,
\label{actEH}
\ee
where we have introduced the coordinate $x^0=ct$. Due to the variations to $N$,
$N_i$ and $g_{ij}$, one can get the equations of motion. The similar motion equations
can be found in \cite{LuMeiPope}. When $\lambda=1$ and $\Lambda_W=0$, a static,
spherically symmetric black hole solution
that satisfies the full set of motion equations is presented as \cite{Kehagias}
\bea
  ds^2 &=& -f(r) dt^2+\frac{dr^2}{f(r)}+r^2
\left(d\theta^2+\sin^2\theta d\phi^2\right) \, , \nn \\
f(r)&=& \frac{2r^2+2P^2-4rM}{r^2+2P^2+\sqrt{r^4+8P^2Mr}}
\, ,   \label{flatBH}
\eea
where $P^2=\kappa^2/(32\mu^2)$, and $M$ is an integration constant. This black hole
is formally different from the usual Schwarzschild black hole. However, for $r\gg
2(MP^2)^{(1/3)}$, we find that $f(r)$ takes the form
\be
f(r) =1-\frac{2M}{r} +\mathcal{O}\left(\frac{1}{r^4}\right) \, ,
\ee
which implies that the solution (\ref{flatBH}) behaves like the usual Schwarzschild
black hole in general relativity when $r$ is sufficiently large. If
$P\rightarrow 0$ and $r$ is finite, we have $f(r)=1-2M/r$, meaning that Eq.
(\ref{flatBH}) exactly coincides with the Schwarzschild black hole. Its two event
horizons that are very similar with those of the Reissner-Nordstr\"{o}m black hole
are
\be
r_\pm = M\pm\sqrt{M^2-P^2} \, ,
\ee
where $M\geq |P|$ to avoid the naked singularity. $M=|P|$ is
the extremity condition. Some properties of the black hole (\ref{flatBH})
have been investigated in \cite{CiteKeha}.

By virtue of the surface gravity formula, the Hawking temperature of the black hole
(\ref{flatBH}) is given by
\be
T_H=\frac{1}{4\pi}\partial_rf(r)\mid_{r_+}
=\frac{r_+^2 -P^2 }{ 4\pi (r_+^3+2P^2r_+)} \label{SurgravityT} \, .
\ee
Obviously, when $M=|P|$, the temperature for the extremal black hole vanishes.
The ADM mass is $M$. If we naively use the area law to calculate the entropy,
we obtain the evaluation that does not fulfill the first law of black hole
thermodynamics. To keep the differential form of the first law hold, i.e.
$dM =T_HdS$, we have to set the entropy $S =\pi r_+^2 +4\pi P^2 \ln(r_+) +S_0$
\cite{ACast}, where $S_0$ is an arbitrary constant irrelevant with $r_+$.
However, in this case, the integral form of the first law is not satisfied.
Maybe we can modify both the mass and entropy to guarantee that the differential
and integral forms of the fist law hold at the same time. We do not discuss
this point in the present paper. In what follows, we shall focus on reproducing
the temperature (\ref{SurgravityT}) via the methods of gravitational anomaly
cancellation and the effective action, as well as the DRS approach.

\section{Hawking radiation via covariant gravitational anomaly}\label{three}

In this section, we will exploit the Hawking radiation of the black hole
(\ref{flatBH}) via the covariant gravitational anomaly cancellation method
\cite{BKulk} developed due to \cite{RW,IUW}. Before doing this, we first give
a brief review of this method. Considering a complex scalar field in the
background of the black hole, since the field near the horizon can be
effectively described by an infinite collection of (1+1)-dimensional fields,
it is feasible for us to treat the original higher dimensional theories as a
set of scalar fields in the background of (1+1)-dimensional space-time. Near
the horizon of this two dimensional reduced space-time, there exist outgoing
modes and ingoing modes. If we omit the classically irrelevant ingoing modes,
the (1+1)-dimensional effective field theory becomes chiral, which leads to
gravitational anomaly. In order to cancel this anomaly, a compensating current
that is identified to the Hawking flux of energy-momentum arises. It is worth
noting that one has to impose proper boundary conditions that compatible with
the Unruh vacuum to fix the Hawking flux completely.

Now we discuss the dimension reduction by considering the action of a massless
scalar fields in background of the black hole (\ref{flatBH}). Performing the
partial wave decomposition $\varphi = \sum_{lm}\varphi_{lm}(t, r) Y_{lm}(\theta, \phi)$,
where $Y_{lm}$ are the spherical harmonics, and only keeping the dominant terms
near the horizon,
\bea
S[\varphi] &=& -\frac{1}{2}\int d^{4}x\sqrt{-g}
g^{\mu\nu}\p_{\mu}\varphi\p_{\nu}\varphi \nn \\
&=& \frac{1}{2}\int dtdrd\theta d\phi\sin\theta \varphi\Big\{-\frac{r^2}{f}\p_t^2
+\p_r(r^2f\p_r) +\Delta_{\Omega} \Big\}\varphi \nn \\
 &=& \frac{1}{2}\sum_{lm}\int dtdr (r^2)\varphi_{lm}
\Big\{-\frac{1}{f}\p_t^2
+\p_r(f\p_r)\Big\}\varphi_{lm} \, ,
\eea
where $\Delta_{\Omega}$ is the angular Laplace operator, and the last equation is
obtained by taking the near horizon limit. The above equation implies that the
scalar field in the original (3+1)-dimensional space-time can be effectively
described by an infinite set of two dimensional fields in the background of the
metric
\be
ds^2 = -f(r)dt^2 + \frac{dr^2}{f(r)} \, ,
\label{effemetric}
\ee
together with the dilaton field $\Psi=r^2$, whose contribution can be neglected,
since the reference space-time is static. Obviously, the effective metric
(\ref{effemetric}) is just the $(t,r)$ part of the original metric (\ref{flatBH}).

Near the horizon, if we omit the quantum effect of the ingoing modes, the effective
field theory will exhibit a gravitational anomaly. For the right-handed fields, the
($1 + 1$)-dimensional covariant gravitational anomaly has the form
\be
 \nabla_{\mu}T_{~\nu}^{\mu} = \frac{-1}{96\pi}\sqrt{-g_2}\epsilon_{\mu\nu}\p^{\mu}R
 = \frac{1}{\sqrt{-g_2}}\p_{\mu}N_{~\nu}^{\mu}  \, , \label{anomalyEM}
\ee
where $\epsilon_{\mu\nu}$ is the two dimensional tensor density with
$\epsilon^{rt}=\epsilon_{tr} =-1$, $R$ is the scalar curvature of the two dimensional
space-time, and $g_2$ is the determinant of the two dimensional metric. In terms of
the effective metric (\ref{effemetric}), the anomaly
is time-like, i.e. $\nabla_{\mu}T_{~r}^{\mu}=0$, and
\be
 N_{~t}^{r} = \frac{1}{192\pi}\big(2ff^{\prime\prime}
  -f^{\prime 2}\big) \, ,
\ee
where and in what follows, the prime denotes the derivative to the radial coordinate.
Introducing two scalar functions $\Theta(r) = \Theta(r -r_+ -\varepsilon)$ and
$H(r) = 1 -\Theta(r)$, where the limit $\varepsilon \to 0$ will be taken
ultimately, the total energy-momentum tensor can be written as
\be
T_{~\nu}^{\mu} = T_{(O)\nu}^{\mu}\Theta(r)
+T_{(H)\nu}^{\mu}H(r) \, ,
\ee
in which $T_{(O)\nu}^{\mu}$, localized in the region outside the horizon
$(r>r_+ +\varepsilon)$, is covariantly conserved, while $T_{(H)\nu}^{\mu}$
that is defined in the near-horizon region $(r_+< r<r_+ +\varepsilon)$
satisfies the anomalous Eq. (\ref{anomalyEM}). Solving both the equations,
we have
\begin{subequations}
\begin{align}
T_{(O)t}^r &= a_O \, , \\
T_{(H)t}^r &= a_H +N_{~t}^r(r) -N_{~t}^r(r_+) \, ,
\end{align}
\end{subequations}
where $a_O$ and $a_H$ are two integration constants, and $a_O$ is the value
of the energy flow at infinity. In order to fix it completely, firstly we
take into account the $\nu = t$ component of Eq. (\ref{anomalyEM}), which
reads
\be
\nabla_{\mu}T_{~t}^{\mu} = \p_r T_{~t}^r
= \p_r\big[N_{~t}^rH(r)\big]
 +\big[T_{(O)t}^r - T_{(H)t}^r
 +N_{~t}^r\big]\delta(r -r_+) \, .
\ee
The first term in the above equation should be cancelled by the ingoing modes
. For the sake of keeping the energy-momentum tensor anomaly free, the second
term must vanish at the horizon. Therefore, we obtain
\be
a_O = a_H- N_{~t}^r(r_+) \, .
\ee
However, this equation is not sufficient to fix energy flow $a_O$ completely.
One has to impose a regular boundary condition corresponding to the Unruh
vacuum. Such a constraint requires the covariant energy-momentum tensor
$T_{~\nu}^{\mu}$ to vanish at the horizon, which yields $a_H = 0$. Thus the
total energy flux at infinity is read off as
\be
a_O = -N_{~t}^r(r_+)
= \frac{f^{\prime 2}(r_+)}{192\pi}
= \frac{1}{192 \pi}\left(\frac{r_+^2 -P^2 }{r_+^3+2P^2 r_+}\right)^2
\, .
\label{energyflow}
\ee
The energy fluxes (\ref{energyflow}), obtained via the covariant gravitational
anomaly, is compatible with the Hawking ones. To see this, we calculate the
Hawking fluxes by considering the fermionic Plack distribution $N(\omega)=
1/(e^{\omega/T_H}+1)$ for blackbody radiation in the general spherically
symmetric black hole background. With such a distribution, the Hawking fluxes
of the energy-momentum tensor is computed as
\be
F_M=\int_0^\infty \frac{\omega}{\pi(e^{\omega/T_H}+1)} d\omega
 =\frac{\pi}{12}T_H^2 \, . \label{Hawkingflu}
\ee
Substituting the Hawking temperature (\ref{SurgravityT}) into Eq.
(\ref{Hawkingflu}), we have $a_O=F_M$. This implies that we can also derive
the same temperature as Eq. (\ref{SurgravityT}) via the method of the covariant
gravitational anomaly cancellation.

\section{Hawking radiation through the method of covariant effective action}\label{four}

It has been shown that the quantum field can be effectively described by an infinite
collection of (1+1)-dimensional fields with the metric (\ref{effemetric}) in the
near-horizon region of the black hole (\ref{flatBH}). Thus we shall employ the
covariant effective action method \cite{BanerKeff} to calculate the Hawking fluxes
in terms of the background of the effective metric (\ref{effemetric}) in this section.
Our calculations only involve the gravitational part of the effective action since
there only exists gravitational anomaly near the horizon. Some results can be found
in our work \cite{Wueff}. Due to the effective action method, the Hawking fluxes are
associated with the $(r,t)$-component of the covariant anomalous energy-momentum
tensor near the horizon, which can be derived by varying the effective action. With
help of the boundary condition that the covariant energy-momentum tensor vanishes
at the horizon, one obtains the Hawking flux that is equal to the $(r,t)$-component
of the energy-momentum tensor at infinity ($r\rightarrow\infty$).

In the (1+1)-dimensional black hole background, the covariant energy momentum tensor
is given by
\cite{HLca}
\be
T_{~\nu}^{\mu} = -\frac{1}{192\pi}\left(D^{\mu}\psi D_{\nu}\psi
 -2D^{\mu}D_{\nu}\psi +2\delta^{\mu}_{~\nu}R\right) \, .
\label{coEMeff}
\ee
For Eq. (\ref{coEMeff}), the auxiliary field $\psi$ obeys the equation
\be
\nabla_{\mu}\nabla^{\mu}\psi = R =-f^{\prime\prime} \, ,
\label{arxiequ}
\ee
which vanishes at infinity, while the chiral covariant differential operator
$D_{\mu}$ is defined by
\be
D_{\mu} = \nabla_{\mu} +\sqrt{-g}\epsilon_{\mu\nu}\nabla^{\nu}
 = \sqrt{-g}\epsilon_{\mu\nu}D^{\nu} \, .
\ee
One can verify that the covariant energy-momentum tensor (\ref{coEMeff}) satisfies
both the covariant anomaly equation (\ref{anomalyEM}) and the conformal trace
anomaly equation
\be
T_{~\mu}^\mu = -\frac{R}{48\pi} \, .
\ee

For the two-dimensional effective metric (\ref{effemetric}), the equation of motion
(\ref{arxiequ}) for the auxiliary field $\psi$ becomes
\be
-\frac{1}{f}\p_t^2\psi +\p_r\big(f\p_r\psi\big) = -\p^2_r f \, .
\ee
The general solution of this equation is
\begin{subequations}
\begin{align}
\psi &= at +\int \frac{b -f^\prime}{f} dr = at  -\ln (f)+ b\int \frac{dr}{f}  \, , \\
D_t\psi &= -D^r\psi = a -b +f^\prime \, ,
\end{align}
\end{subequations}
where $a$ and $b$ are constants to be specified by imposing appropriate boundary
condition. Therefore, the $(r,t)$-component of the covariant energy momentum
tensor (\ref{coEMeff}) can be computed as
\be
T_{~t}^{r} = \frac{1}{192\pi}
 \big[(a -b)^2 -f^{\prime 2} +2ff^{\prime\prime}\big] \, .
\label{EManomH}
\ee
As before, we still choose the boundary condition that the covariant energy
momentum tensor vanishes at event horizon. Thus we have
\be
(a-b)^2= f^{\prime 2}(r_+) \, .
\label{relaofab}
\ee
In \cite{BanerKeff}, it has been shown that the energy flux is given by the
asymptotical limit of the anomaly free energy momentum tensor. For the
effective metric (\ref{effemetric}), since $f^\prime(\infty) =0$  and
$f^{\prime\prime}(\infty) =0$, the Eq. (\ref{anomalyEM}) becomes
$\nabla_\mu T^\mu_{~\nu}\rightarrow0$ when $r$ goes to infinity,
which implies that the energy momentum tensor is anomaly free at infinity.
Therefore, the energy flux reads
\be
T^r_{~t}(r\rightarrow\infty)=\frac{ f^{\prime 2}(r_+)}{192\pi}
=\frac{1}{192 \pi}\left(\frac{r_+^2 -P^2 }{r_+^3+2P^2 r_+}
\right)^2 \, .
\label{enereff}
\ee
This energy flux agrees with Eq. (\ref{energyflow}) got via the method of
covariant anomaly cancellation.

Besides, we can also evaluate the energy flux by calculating the anomaly
free energy momentum tensor outside the horizon, which reads \cite{HLca}
\be
T^\mu_{(O)\nu}=-\frac{1}{96\pi}\left(4\delta^\mu_{~\nu}R-4\nabla^\mu\nabla_\nu\psi
+2\nabla^\mu\psi\nabla_\nu\psi -\delta^\mu_{~\nu}
\nabla^\sigma\psi\nabla_\sigma\psi \right) \, .
\ee
This anomaly-free energy momentum satisfies $\nabla_\mu T^\mu_{(O)\nu}=0$.
Its $(r,t)$ component is
\be
T^r_{(O)t} = -\frac{ab}{48\pi} \, .
\label{EMfluxOut}
\ee
Here we do not give the other components of the energy momentum tensor
$T^\mu_{(O)\nu}$. All the components coincide with their corresponding
ones obtained via the Unruh vacuum approach \cite{RotatingAnom} if we
still choose the boundary condition that the covariant energy momentum
disappears at the horizon. Since Eq. (\ref{anomalyEM}) vanishes at
infinity, Eq. (\ref{EMfluxOut}) must be in agreement with Eq.
(\ref{EManomH}) when we take the asymptotical limit. Thus we have
$a =-b$. With the help of Eq. (\ref{relaofab}), we obtain
\begin{subequations}
\begin{align}
a &= f^{\prime}(r_+)/2 \, , \qquad b = -f^{\prime}(r_+)/2 \, , \\
a &= -f^{\prime}(r_+)/2 \, , \qquad b = f^{\prime}(r_+)/2 \, .
\end{align}
\end{subequations}
Substituting each of these equations into Eq. (\ref{EMfluxOut}), we get
the energy flux that takes the same form as Eq. (\ref{enereff}).
Looking through the whole procedure, we find that the asymptotical behaviour
of $f(r)$ plays a key role in our derivation. If $f^{\prime 2}
-2ff^{\prime\prime}$ is divergent when $r\rightarrow\infty$, all of the
derivation is invalid. For example, our calculation in this section
fails to give the energy flux of the Schwarzschild AdS black hole. Finally,
from the energy flux, we can obtain the Hawking temperature in agreement
with Eq. (\ref{SurgravityT}).

\section{Hawking radiation via the DRS approach}\label{five}

In this section, we will make use of the DRS approach \cite{DRS} to investigate
Hawking radiation of the black hole (\ref{flatBH}) in terms of the (1+1)-dimensional
effective metric (\ref{effemetric}). This approach has also been employed to study
Hawking radiation from the Reissner-Nordstr\"{o}m black hole with a global monopole
in \cite{WuPCQG1}. We now take into account the Klein Gordon equation of a complex
scalar field $\Phi$ with mass $\mu$, which reads
\be
-\frac{1}{f}\p_t^2\Phi +\p_r\big(f\p_r\Phi\big) -\mu^2 \Phi = 0 \, .
\ee
Separating the scalar field $\Phi$ as $\Phi(t,r) =R(r)e^{-i\omega t}$ and performing
the coordinate transformation $r_\ast =\int dr/f$, the radial equation is given by
\be
\frac{d^2R(r)}{dr_\ast^2}+(\omega^2-\mu^2f) R(r)=0  \, .
\ee
Near the horizon, $f(r)\rightarrow 0$. Thus the scalar field $\Phi$ can be solved as
\begin{subequations}
\begin{align}
\Phi_{in} &= e^{-\mathrm{i}\omega (t +r_\ast)} \, , \\
\Phi_{out} &= e^{-\mathrm{i}\omega (t -r_\ast)}
 =\Phi_{in}e^{2\mathrm{i}\omega r_\ast}\, ,
\end{align}
\end{subequations}
where $\Phi_{in}$ is the ingoing wave solution, which is analytic at the horizon,
while  $\Phi_{out}$, the outgoing wave solution, is logarithmically singular at
the horizon. To see this, by virtue of the relation $r_\ast \approx
\ln(r-r_+)/(2\kappa)$ near the horizon, where $\kappa =f^\prime(r_+)/2$ is the
surface gravity of the black hole (\ref{flatBH}), we can reexpress the outgoing
wave solution as
\be
\Phi_{out}\approx \Phi_{in}(r-r_+)^{\mathrm{i}\omega/\kappa} \qquad
(r>r_+) \, .
\ee
However, it can be analytically continued to
\be
\widetilde{\Phi}_{out} = \Phi_{in}(r_+ -r)^{\mathrm{i}\omega/\kappa}
e^{\pi \omega / \kappa} \qquad  (r<r_+)
\ee
from the outside of the hole into the inside hole along the lower $r$-plane.
Therefore, we obtain the relative scattering probability at the horizon
\be
\left|\frac{\Phi_{\rm out}}{\widetilde{\Phi}_{\rm out}}\right|^2
= e^{-2\pi\omega /\kappa} \, .
\ee
Following the DRS approach proposed in \cite{DRS}, the thermal spectrum of the
particles radiating from the black hole can be given by
$N(\omega)=1/(e^{2\pi\omega/\kappa}- 1)$
for scalar particles. From the thermal spectrum, we can obtain the Hawking
temperature $T_H =f^\prime(r_+)/(4\pi)$, which takes the same form as
Eq. (\ref{SurgravityT}).

\section{Hawking radiation of black holes with cosmological constant}\label{six}

In this section, we shall extend the gravitational anomaly cancellation method to
exploit the Hawking radiation of the static, spherically symmetric black hole
with cosmological constant in the IR Modified HL  Gravity.
This black hole solution, which fulfills the motion equations generated from
the action (\ref{action}) with the coupling constant $\lambda=1$, is given
by \cite{MIPark}
\bea
ds^2 &=& -h(r) dt^2+\frac{dr^2}{h(r)}+r^2
\left(d\theta^2+\sin^2\theta d\phi^2\right) \, , \nn \\
h(r)&=& \frac{1}{2P^2}\Big[2P^2+r^2(1-2\Lambda_WP^2)
-\sqrt{r^4(1-4\Lambda_WP^2)+4rmP^4} \Big]
\, ,   \label{blackhole}
\eea
where $m$ is an integration constant. Choosing proper
values of $\Lambda_W$, $P$ and $m$, the solution (\ref{blackhole}) can recover
some known solutions. We first set $\Lambda_W=0$ and $m=2M/ P^2$. In such a case,
the solution (\ref{blackhole}) becomes the one (\ref{flatBH}), which agrees
with the usual Schwarzschild black hole. When $\Lambda_W$ $ <0$,  for $r \gg [mP^4 /
( 1-4 P^2\Lambda_W)]^{1/3}$ and $m=2M/ P^2$, Eq. (\ref{blackhole}) reduces
to the standard Schwarzschild-AdS black hole except for a factor ``$|2\Lambda_WP^2|$"
\cite{MIPark}. Besides, in this case, if $P\rightarrow\infty$ and $m=\alpha^2
\sqrt{-\Lambda_W}$,
Eq. (\ref{blackhole}) takes the same form as that in
\cite{LuMeiPope}. Finally, by setting $\Lambda_W>0$ and performing the transformation
$\mu \rightarrow i \mu$, $w^2 \rightarrow -i w^2$, one can obtain the usual
Schwarzschild-dS-like solution.

Since the analysis via the anomaly cancellation approach here is parallel with what we
have performed in the section (\ref{three}), we only present some main results. By the
dimension reduction technique, we have the two dimensional effective metric
\be
ds^2 = -h(r)dt^2 + \frac{dr^2}{h(r)} \, .
\ee
In terms of this effective metric, the Hawking flux can be computed as
\be
a_O = \frac{h^{\prime 2}(r_+)}{192\pi}
= \frac{1}{192 \pi}\left\{\frac{3 \Lambda_W^2P^2 r_+^4
+(1-2P^2\Lambda_W)r_+^2 -P^2 }
{r_+ [2P^2 +(1-2P^2\Lambda_W)r_+^2]}\right\}^2  \, ,
\ee
where $r_+$, the largest root of the equation $h(r)=0$, is the outside horizon
of the black hole (\ref{blackhole}). From the Hawking flux, we obtain the
Hawking temperature, which reads
\be
T_H =\frac{3 \Lambda_W^2P^2 r_+^4 +(1-2P^2\Lambda_W)r_+^2 -P^2 }
{ 4 \pi r_+ [2P^2 +(1-2P^2\Lambda_W)r_+^2]} \label{SurgT} \, .
\ee
This temperature coincides with the one derived by virtue of the surface gravity formula.
In particular, when $\Lambda_W =0$, it becomes that of the asymptotically flat black hole
(\ref{flatBH}). Besides, we can also derive the same temperature via the DRS method
following the above section.

\section{Summary}\label{seven}

In this paper, we have extended the methods of covariant anomaly cancellation and the
effective action to derive the Hawking temperature of the black holes \cite{Kehagias}
in the IR modified Ho\v{r}ava-Lifshitz gravity. These black holes are the spherically
symmetric, asymptotically flat solutions when the coupling constant $\lambda=1$, namely,
the HL gravity returns to general relativity. They are formally
different from the usual Schwarzschild black holes but agree with the Schwarzschild
ones when the radial distance is sufficiently large. The crucial points of both the
two methods are the dimension reduction and the appropriate boundary condition. Using
the dimension reduction technique that the complex scalar field in the background of
the higher dimensional spacetime can be effectively described by an infinite
collection of two dimensional scalar fields, we obtained the two dimensional effective
metric. On basis of the effective metric, imposing the boundary condition that the
fluxes of the energy momentum tensor vanishes at the event horizon, we applied both
the methods to calculate the Hawking fluxes. The values obtained via these two methods
are equal. Our results further support Hawking radiation is a universal quantum
behavior arising at the event horizon. To see the universality of Hawking radiation,
the DRS method have also been used to compute the Hawking temperature in terms of the
two dimensional effective metric.

The covariant anomaly cancellation method has also been generalized to study the
Hawking radiation of the spherically symmetric, asymptotically AdS black holes
\cite{MIPark}. These black holes can be seen as the analogs of the usual
Schwarzschild AdS black holes. We have obtained their Hawking temperature which
recovers that of the asymptotically flat black holes in \cite{Kehagias} when
$\Lambda_W =0$.

Finally, we give a simple comment on the methods of anomaly cancellation and effective
action. Both the methods are applicable to the theories that are generally covariant.
Although the HL gravity theory is invariant under foliation preserving diffeomorphism,
we can still use the two methods to study Hawking radiation of black holes
in this gravity theory, since the group with foliation preserving diffeomorphism is
a subgroup of the one with full diffeomporphism. Further investigation is required
to prove the strict validity of applying both the methods to the gravity theory
with foliation preserving invariance.

\section*{Acknowledgments}

This work was partially supported by the Natural Science Foundation of China under
Grant Nos. 10975058 and 10675051. J.J. Peng was also supported in part by a Graduate
Innovation Foundation of Central China Normal University.


\begin{thebibliography}{99}

\bibitem{Horava}
P. Ho\v{r}ava, Phys. Rev. D \textbf{79}, 084008 (2009),
arXiv:0901.3775 [hep-th];
%
P. Horava, J. High Energy Phys. \textbf{03}, 020 (2009),
arXiv:0812.4287 [hep-th].

\bibitem{Kehagias}
A. Kehagias, K. Sfetsos, Phys. Lett. B \textbf{678}, 123 (2009),
arXiv:0905.0477 [hep-th].

\bibitem{LuMeiPope}
H. Lu, J.W. Mei, C.N. Pope, Phys. Rev. Lett. \textbf{103}, 091301 (2009),
arXiv:0904.1595 [hep-th].

\bibitem{MIPark}
M.I. Park, J. High Energy Phys. \textbf{09}, 123 (2009), arXiv:0905.4480 [hep-th].

\bibitem{Solutions}
A. Ghodsi, E. Hatefi, arXiv:0906.1237 [hep-th];
%
M. Botta-Cantcheff, N. Grandi, M. Sturla, arXiv:0906.0582 [hep-th];
%
A. Ghodsi, arXiv:0905.0836 [hep-th];
%
E.\'{O}. Colg\'{a}in, H. Yavartanoo, J. High Energy Phys. \textbf{08}, 021 (2009), arXiv:0904.4357 [hep-th];
%
R.G. Cai, L.M. Cao, N. Ohta, Phys. Rev. D \textbf{80}, 024003 (2009), arXiv:0904.3670 [hep-th];
%
H. Nastase, arXiv:0904.3604 [gr-qc].

\bibitem{CiteKeha}
Y.W. Kim, H.W. Lee, Y.S. Myung, Phys. Lett. B \textbf{682}, 246 (2009),
arXiv:0905.3423 [hep-th];
%
J.H. Chen, Y.J. Wang, arXiv:0905.2786 [gr-qc];
%
S.B. Chen, J.L. Jing, Phys. Rev. D \textbf{80}, 024036 (2009), arXiv:0905.2055 [gr-qc];
%
S.B. Chen, J.L. Jing, arXiv:0905.1409 [gr-qc];
%
Y.S. Myung, arXiv:0906.0848 [hep-th];
%
Y.S. Myung, Phys. Lett. B \textbf{678}, 127 (2009), arXiv:0905.0957 [hep-th].

\bibitem{ACast}
A. Castillo, A. Larranaga, arXiv:0906.4380 [gr-qc].

\bibitem{BHphy}
R.G. Cai, L.M. Cao, N. Ohta, Phys. Lett. B \textbf{679}, 504 (2009), arXiv:0905.0751 [hep-th];
%
Y.S. Myung, Y.W. Kim, arXiv:0905.0179 [hep-th].

\bibitem{quafiled}
F.W. Shu, Y.S. Wu, arXiv:0906.1645 [hep-th];
%
G. Calcagni, arXiv:0905.3740 [hep-th];
%
C. Charmousis, G. Niz, A. Padilla, P.M. Saffin, J. High Energy Phys. \textbf{08}, 070 (2009),
arXiv:0905.2579 [hep-th];
%
M. Li, Y. Pang, J. High Energy Phys. \textbf{08}, 015 (2009), arXiv:0905.2751 [hep-th];
%
D. Orlando, S. Reffert, Class. Quantum Grav. \textbf{26}, 155021 (2009),
arXiv:0905.0301 [hep-th];
%
B.Chen, Q.G. Huang, Phys. Lett. B (in press), arXiv:0904.4565 [hep-th];
%
T.P. Sotiriou, M. Visser, S. Weinfurtner, JHEP \textbf{10}, 033 (2009),
arXiv:0905.2798 [hep-th];
%
S. Mukohyama, JCAP \textbf{09}, 005 (2009),
arXiv:0906.5069 [hep-th];
%
J. Kluson, JHEP \textbf{07}, 079 (2009),
arXiv:0904.1343 [hep-th].

\bibitem{gwave}
T. Takahashi, J. Soda,  Phys. Rev. Lett. \textbf{102}, 231301 (2009),
arXiv:0904.0554 [hep-th].

\bibitem{cosmolo}
A.Z. Wang, Y.M. Wu, JCAP \textbf{07}, 012 (2009), arXiv:0905.4117 [hep-th];
%
X. Gao, Y. Wang, R. Brandenberger, A. Riotto, arXiv:0905.3821 [hep-th];
%
C.J. Gao, arXiv:0905.0310 [astro-ph.CO];
%
Y.S. Piao, arXiv:0904.4117 [hep-th];
%
G. Calcagni, JHEP \textbf{09}, 112 (2009), arXiv:0904.0829 [hep-th].

\bibitem{SWH}
S. Hawking, Nature (London) \textbf{248}, 30 (1974);
%
Commun. Math. Phys. \textbf{43}, 199 (1975).

\bibitem{RW}
S.P. Robinson, F. Wilczek, Phys. Rev. Lett. \textbf{95}, 011303 (2005), gr-qc/0502074.

\bibitem{IUW}
S. Iso, H. Umetsu, F. Wilczek, Phys. Rev. Lett. \textbf{96}, 151302 (2006), hep-th/0602146.

\bibitem{RotatingAnom}
S. Iso, H. Umetsu, Frank Wilczek, Phys. Rev. D \textbf{74}, 044017 (2006), hep-th/0606018.

\bibitem{MurataJ}
K. Murata, J. Soda, Phys. Rev. D \textbf{74}, 044018 (2006), hep-th/0606069.

\bibitem{RotAnomB}
S. Iso, T. Morita, H. Umetsu, JHEP \textbf{04}, 068 (2007), hep-th/0612286;
%
Z.B. Xu, B. Chen, Phys. Rev. D \textbf{75}, 024041 (2007), hep-th/0612261;
%
H. Shin, W. Kim, J. High Energy Phys. \textbf{06}, 012 (2007), arXiv:0705.0265 [hep-th];
%
W. Kim, H. Shin, J. High Energy Phys. \textbf{07}, 070 (2007), arXiv:0706.3563 [hep-th];
%
A.P. Porfyriadis, Phys. Rev. D \textbf{79}, 084039 (2009), arXiv:0811.2822 [hep-th];
%
A.P. Porfyriadis, Phys. Lett. \textbf{B} 675, 235 (2009), arXiv:0904.2042 [hep-th].

\bibitem{BonoraC}
L. Bonora, M. Cvitan, J. High Energy Phys. \textbf{05}, 071 (2008), arXiv:0804.0198 [hep-th];
%
L. Bonora, M. Cvitan, S. Pallua, I. Smoli\'{c}, J. High Energy Phys. \textbf{12}, 021 (2008),
arXiv:0808.2360 [hep-th].

\bibitem{WuPCQG1}
S.Q. Wu, J.J. Peng, Class. Quantum Grav. \textbf{24}, 5123 (2007), arXiv:0706.0983 [hep-th]

\bibitem{Sphano}
J.J. Peng, S.Q. Wu, Chin. Phys. B \textbf{17}, 825 (2008), arXiv:0705.1225 [hep-th];
%
E. Papantonopoulos, P. Skamagoulis, Phys. Rev. D \textbf{79}, 084022 (2009), arXiv:0812.1759 [hep-th];
%
S.W. Wei, R. Li, Y.X. Liu, J.R. Ren, Eur. Phys. J. C \textbf{65}, 281 (2010), arXiv:0901.2614 [hep-th].

\bibitem{Wueff}
S.Q. Wu, J.J. Peng, Z.Y. Zhao, Class. Quantum Grav. \textbf{25}, 135001 (2008), arXiv:0803.1338 [hep-th].

\bibitem{BRing}
U. Miyamoto, K. Murata, Phys. Rev. D \textbf{77}, 024020 (2008), arXiv:0705.3150 [hep-th];
%
B. Chen, W. He, Class. Quantum Grav. \textbf{25}, 135011 (2008), arXiv:0705.2984 [gr-qc].

\bibitem{BKulk}
R. Banerjee, S. Kulkarni, Phys. Rev. D \textbf{77}, 024018 (2008), arXiv:0707.2449 [hep-th].


\bibitem{CAstr}
J.J. Peng, S.Q. Wu, Phys. Lett. B \textbf{661}, 300 (2008), arXiv:0801.0185 [hep-th].

\bibitem{GangKul}
S. Gangopadhyay, S. Kulkarni, Phys. Rev. D \textbf{77}, 024038 (2008), arXiv:0710.0974 [hep-th];
%
S. Nam, J.D. Park, Class. Quantum Grav. \textbf{26}, 145015 (2009), arXiv:0902.0982 [hep-th];
%
J.J. Peng, S.Q. Wu, Gen. Rel. Grav. \textbf{40}, 2619 (2008), arXiv:0709.0167 [hep-th].

\bibitem{Baner}
R. Banerjee, Int. J. Mod. Phys. D \textbf{17}, 2539 (2009), arXiv:0807.4637 [hep-th];
%
R. Banerjee, B.R. Majhi, Phys. Rev. D \textbf{79}, 064024 (2009), arXiv:0812.0497 [hep-th].




\bibitem{BanerKeff}
R. Banerjee, S. Kulkarni, Phys. Lett. B \textbf{659}, 827 (2008), arXiv:0709.3916 [hep-th];
%
R. Banerjee, S. Kulkarni, Phys. Rev. D \textbf{79}, 084035 (2009), arXiv:0810.5683 [hep-th].

\bibitem{HLca}
H. Leutwyler, Phys. Lett. B \textbf{153}, 65 (1985); \textbf{155}, 469(E) (1985).

\bibitem{AShirasakaT}
A. Shirasaka, T. Hirata, arXiv:0804.1910 [hep-th].

\bibitem{effappli}
S. Gangopadhyay, Phys. Rev. D \textbf{77}, 064027 (2008), arXiv:0712.3095 [hep-th];
%
S. Kulkarni, Class. Quantum Grav. \textbf{25}, 225023 (2008), arXiv:0802.2456 [hep-th].

\bibitem{DRS}
T. Damour, R. Ruffini, Phys. Rev. D \textbf{14}, 332 (1976);
S. Sannan, Gen. Rel. Grav. \textbf{20}, 239 (1988).


\end{thebibliography}
\end{document}